%% file: sample-sigconf.tex
\documentclass[sigconf]{acmart}

\usepackage{booktabs} 
\usepackage{flushend}

\setcopyright{acmcopyright}

\acmDOI{10.475/123_4}

\acmISBN{123-4567-24-567/08/06}

\acmConference[Submitted to KDD'18]{}{August 19--23}{London, United Kingdom}
\acmYear{2018}
\copyrightyear{2018}

\acmPrice{15.00}

\begin{document}
\title{Wikipedia graph mining: dynamic structure of collective memory}

\author{Volodymyr Miz, Kirell Benzi, Benjamin Ricaud, and Pierre Vandergheynst}
\affiliation{%
  \institution{Ecole Polytechnique F\'ed\'erale de Lausanne}
  \city{Lausanne}
  \state{Switzerland}
}
\email{first.last@epfl.ch}

\renewcommand{\shortauthors}{V. Miz et al.}

\begin{abstract}
Wikipedia is the biggest encyclopedia ever created and the fifth most visited website in the world. Tens of millions of people surf it every day, seeking answers to various questions. Collective user activity on its pages leaves publicly available footprints of human behavior, making Wikipedia an excellent source for analysis of collective behavior.

In this work, we propose a new method to analyze and retrieve collective memories, the way social groups remember and recall the past. We use the Hopfield network model as an artificial memory abstraction to build a macroscopic collective memory model. To reveal memory patterns, we analyze the dynamics of visitors activity on Wikipedia and its Web network structure. Each pattern in the Hopfield network is a cluster of Wikipedia pages sharing a common topic and describing an event that triggered human curiosity during a finite period of time. We show that these memories can be remembered with good precision using a memory network. The presented approach is scalable and we provide a distributed implementation of the algorithm.
\end{abstract}

%
%



\maketitle

\input{samplebody-conf}
\bibliographystyle{ACM-Reference-Format}
\bibliography{sample-bibliography}

\end{document}

%% file: samplebody-conf.tex
\section{Introduction}\label{Introduction}

Over recent years, the Web has significantly affected the way people learn, interact in social groups, store and share information. Apart from being an essential part of modern life, social networks, online services, and knowledge bases generate a massive amount of logs, containing traces of global online activity on the Web. A large-scale example of such publicly available information is the Wikipedia knowledge base and its history of visitors activity. This data is a great source for collective human behavior analysis at scale. Due to this reason, the analysis of the Wikipedia in this area has become popular over the recent years~\cite{tinati2016finding}, \cite{garcia2017memory}, \cite{kanhabua2014triggers}.

Collective memory \cite{halbwachs2013cadres} is an interesting social phenomenon of human behavior. Studying this concept is a way to enhance our understanding of a common view of events in social groups and identify the events that influence remembering of the past. Early research on collective memory relied on interviews and self-reports that led to a limited number of subjects and biased results~\cite{stone1999science}. The availability of the Web activity data opened new opportunities toward systematic studies at a much larger scale~\cite{ferron2012collective}, \cite{graus2017birth}, \cite{garcia2017memory}. Nonetheless, the general nature of collective memory formation and its modeling remain open questions. Can we model collective and individual memory formation similarly? Is it possible to find collective memories and behavior patterns inside a collaborative knowledge base? In this work, we adopt a data-driven approach to shed some light on these questions.

An essential part of the Web visitors activity data is the underlying human-made graph structure that was initially introduced to facilitate navigation. The combination of the activity dynamics and structure of Web graphs inspires an idea of their similarity to biological neural networks. A good example of such network is the human brain. Numerous neurons in the brain constitute a biological dynamic neural network, where dynamics are expressed in terms of neural spikes. This network is in charge of perception, decision making, storing memories, and learning. 

During learning, neurons in our brain self-organize and form strongly connected groups called neural assemblies~\cite{allport1985distributed}. These groups express similar activation patterns in response to a specific stimuli. When learning is completed, and the stimuli applied once again, reactions of the assemblies correspond to consistent dynamic activity patterns, i.e. memories. Synaptic plasticity mechanisms govern this self-organization process. 

Hebbian learning theory~\cite{hebb2005organization} proposes an explanation of this self-organization and describes the basic rules that guide the network design. The theory implies that simultaneous activation of a pair of neurons leads to an increase in the strength of their connection. In general,the Hebbian theory implies that the neural activity transforms brain networks. This assumption leads to an interesting question. Can temporal dynamics cause a self-organization process in the Wikipedia Web network, similar to the one, driven by neural spikes in the brain?

To answer this question, we introduce a collective memory modeling approach inspired by the theory of learning in the brain, in particular, a content-addressable memory system, the Hopfield network \cite{hopfield1982neural}. In our experiments, we use a dynamic network of Wikipedia articles, where the dynamics comes from the visitors activity on each page, i.e. the number of visits per page per hour. We assume that the Wikipedia network can self-organize similar to a Hopfield network with a modified Hebbian learning rule and learn collective memory patterns under the influence of visitors activity similar to neurons in the brain. The results of our experiments demonstrate that memorized patterns correspond to groups of collective memories, containing clusters of linked pages that have a closely related meaning. A topic of a cluster corresponds to a real world event that triggered the interest of Wikipedia visitors during a finite period of time. In addition, the collective memory gives access to the time lapse when the event occured. We also show that our collective memory model is able to recall an event recorded in the collective memory, given only a part of the event cluster. Here the term \emph{recall} means that we recover a missing fraction of the visitor activity signal in a memory cluster.

\textbf{Contributions.} Our contributions are as follows:
\begin{itemize}
	\item We propose a novel collective memory learning framework, inspired by artificial models of individual memory -- the Hebbian learning theory and the Hopfield network model.
	\item We formalize our findings into an content-addressable model of collective memories. So far, collective memory \cite{halbwachs2013cadres} has been considered just as a concept that lacks a general model of memory formation. In the experiments we demonstrate that given a modified Hebbian learning rule, Wikipedia hyperlinks network can self-organize and gain properties reminiscent of an associative memory.
	\item We develop a graph algorithm for collective memory extraction. Computations are local on the graph, allowing to build an efficient implementation. We provide a distributed implementation of the algorithm to show that it can handle dense graphs and massive amounts of time-series data.
	\item We present graph visualizations as an interactive tool for collective memory studies.
\end{itemize}

The rest of this paper organized as follows. In Section~\ref{related work}, we give an overview of related works on large-scale collective memory research and analysis. Then, we present a graph algorithm and a community detection approach in Section~\ref{Graph learning}. Section~\ref{Dataset} describes the dataset and data preprocessing. We then discuss the results of our experiments and evaluate discovered memory properties in Section~\ref{Experimental results}. Finally, information about the data, the code and tools used for the study as well as online visualizations are provided in Section~\ref{reproducible}.

\section{Related work} \label{related work}

The term \textit{Collective memory} first appeared in the book of Maurice Halbwachs in 1925 \cite{halbwachs2013cadres}. He proposed a concept of a social group that shares a set of collective memories that exist beyond a memory of each member and affects understanding of the past by this social group. Halbwachs's hypothesis influenced a range of studies in sociology \cite{assmann1995collective}, \cite{barash2016collective}, psychology \cite{coman2009collective}, \cite{ferron2012psychological}, cognitive sciences \cite{ferron2012collective}, and, only recently, in computer science \cite{au2011studying}, where authors extract collective memories using LDA \cite{blei2003latent}, applied to a collection of news articles.

Despite the fact that Wikipedia is the largest ever created encyclopedia of public knowledge and the fifth most visited website in the world, the studies on collective memory considered the Wikipedia visitor activity data only recently. The idea of regarding Wikipedia as a global memory space was first introduced by Pentzold in 2009~\cite{pentzold2009fixing}. Then, it was followed by a range of collective memory studies based on various Wikipedia data archives. 

Analyzing Wikipedia page views, Kanhabua et al.~\cite{kanhabua2014triggers} proposed a collective memory model, investigating 5500 events from 11 categories. The authors proposed a remembering score based on a combination of time-series analysis and location information. The focus of the work is on the four types of events: aviation accidents, earthquakes, hurricanes, and terrorist attacks. The work presents extensive experimental results, however, it limited to particular types of collective memories. 

Traumatic events such as attacks and bombings have also been investigated in \cite{ferron2011studying}, \cite{ferron2012collective} based on the Wikipedia edit activity data. The authors investigate the difference between traumatic and nontraumatic events using natural language processing techniques. The study is limited to a certain type of events.

Another case study~\cite{garcia2017memory} focuses on memory-triggering patterns to understand collective memories. The collective memory model is inferred from the Wikipedia visitors activity. The work considers only a case of aircraft incidents reported in English Wikipedia. The authors try to build a general mathematical model and explain the phenomenon of collective memory, extracted from Wikipedia, based on that single topic.

Popularity and celebrities represent another focus point of public interest. The Wikipedia hourly visits on the pages of celebrities was used to investigate fame levels of tennis players \cite{yucesoy2016untangling}. The authors, though, did not tackle collective memories and aimed to quantify the relationship between performance and popularity of the athletes.

To the best of our knowledge, we are the first to apply a content-addressable memory model to the Wikipedia viewership data for collective memory analysis. We build our collective memory model based on the assumption that it is similar to a memory of an individual. For the first time, we demonstrate that Wikipedia Web network and its dynamics can gain properties reminiscent of associative memory. Unlike the results of previous works, our model is not limited to particular classes of events and does not require labeled data. We do not analyze the content of the pages and only rely on the collective viewership dynamics in the network that makes the presented model tractable.

\begin{figure}[t!]
	\centering
	\includegraphics[width=\columnwidth, clip, trim={9.1cm, 6.9cm, 9.2cm, 5.7cm}]{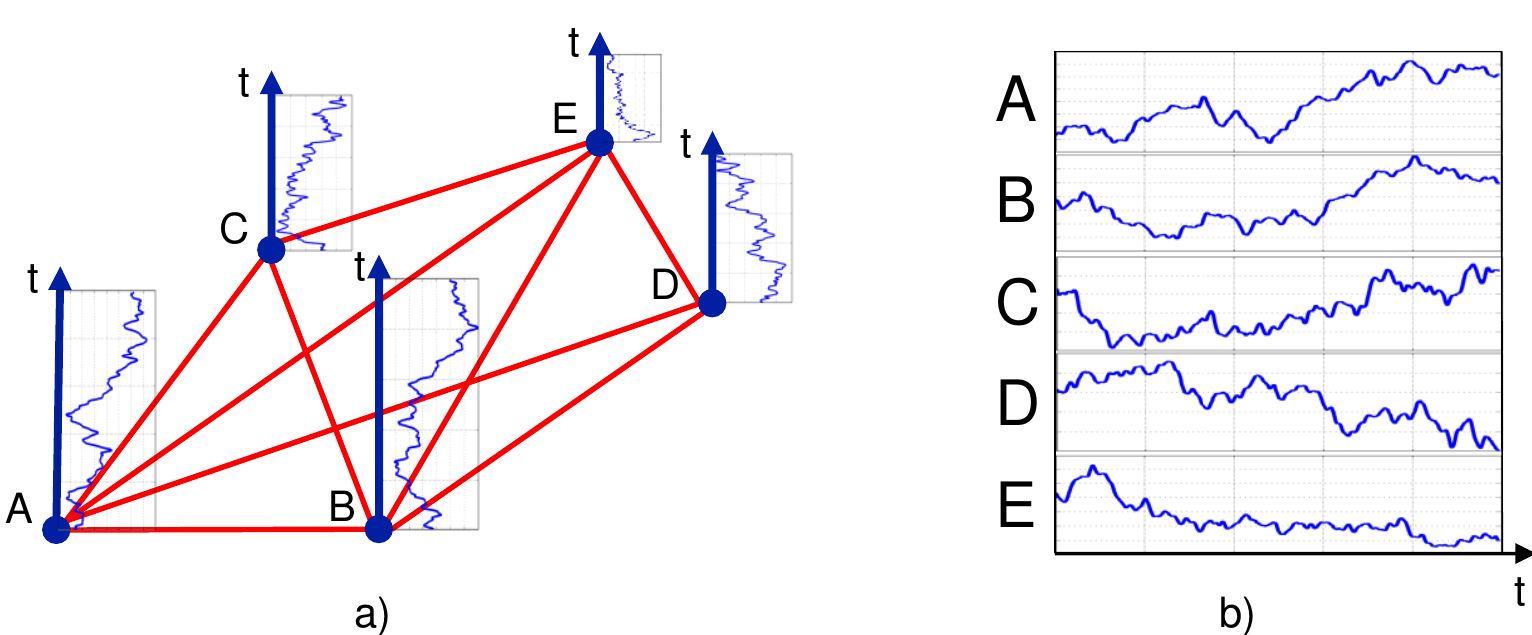}
	\caption{ a) Data structure built from Wikipedia data, with time-series associated to nodes of the Wikipedia hyperlink graph. b) Focus on the time-series signals (number of visits per hour) residing on the vertices of the graph.}
	\label{fig: ts_graph}
\end{figure}

\section{Dataset} \label{Dataset}

We use the dataset described in \cite{benzi2017recommender}.
This dataset is based on two Wikipedia SQL dumps: English language articles and user visit counts per page per hour. The original datasets are publicly available on the Wikimedia website~\cite{wikimediaPageCounts}. 

The Wikipedia network of pages is first constructed using the data from article dumps that contain information about the references (edges) between the pages (nodes)\footnote{Note that Wikipedia is continuously updating. Some links that existed at the moment we made the dump may have been removed from current versions of the pages. To check consistency with past versions, one can use the dedicated search tool at \url{http://wikipedia.ramselehof.de/wikiblame.php}.}. Time-series are then associated to each node (Fig. \ref{fig: ts_graph}), corresponding to the visits history from 02:00, 23 September 2014 to 23:00, 30 April 2015.

The graph contains 116 016 Wikipedia pages (out of a total 4 856 639 pages) with 6 573 475 hyperlinks. Most of the Wikipedia pages remain unvisited. Therefore, only the pages that have a number of visits higher than 500 at least once in the recording are kept. The time-series associated to the nodes have a length of $T=5278$ hours.

\textbf{Preprocessing.} To identify potential core events of collective memories and reduce further the processing time, we select nodes that have bursts of visitor activity, i.e. spikes in the signal. This is done on a monthly basis for two reasons. Firstly, it reduces the processing and the needs for memory as each month can be processed independently. Secondly, it allows for a study of the evolution of the collective memories on a monthly basis, by investigating the trained Hopfield networks (see next section).

For each month, we define the burstiness $b$ of a page as the number and length of the peaks of visits over time. We denote by $x_i$ the time-series giving the number of visits per hour for a node labeled $i$, during month $m$. To define a burst of visits we compute the mean $\mu$ and the standard deviation $\sigma$ of $x_i$. We select values that are above $n\sigma+\mu$, where $n$ is a tunable activity rate parameter ($n=5$ here). The burstiness of the page associated to node $i$ is defined by $ b = \sum_{t = 0}^{T_m-1}k[t] $ where $T_m$ is the length of the time-series for month $m$ and 

\begin{equation} \label{eq: burst}
k[t] = \begin{cases}
1, &\text{if $x_i[t] > n \sigma + \mu$},\\
0, &\text{otherwise}.
\end{cases}
\end{equation}

For each month, we discard the pages that have a burstiness smaller or equal to $5$.




\section{Collective memory learning} \label{Graph learning}

Our goal is to create a self-organized memory from the Wikipedia data. This memory is not a standard neural network but more a concept network or a network made of pieces of information. Indeed, the Wikipedia pages replace neurons in the network. Hence it can be seen as a memory connecting high-level concepts. The self-organized structure is shaped as a Hopfield network with additional constraints, relating it to the concept of associative memory. The learning process follows the Hebbian rules, adapted to our particular dataset.

\textbf{Hopfield network}. Our approach is based on a Hopfield model of artificial memory~\cite{hopfield1982neural}. A Hopfield network is a recurrent neural network serving as an associative memory system. It allows for storing and recalling patterns. Let $N$ be the number of nodes in the Hopfield network. Starting from an initial partial memory pattern $P_0\in \mathbb{R}^N$, the action of recalling a learned pattern is done by the following iterative computation:

\begin{equation} \label{eq: hopfield}
P_{j+1} = f_{\theta}(WP_{j}),
\end{equation}
where $W$ is the weight matrix of the Hopfield network. The function $f_{\theta}:\mathbb{R}^N\to \mathbb{R}^N$ is a nonlinear thresholding function (step function giving values $\{-1,1\}$) that binarize the vector entries. The value $\theta$ is the threshold (same for all entries). In our case, we build a network per month so $W$ is associated to a particular month. For each $j\ge0$, $P_j$ is a matrix of (binarized) time-series where each row is associated to a node of the network and each column corresponds to an hour of the month considered. We stop the iteration when the iterative process has converged to a stable solution ($P_{j+1}=P_j$). Note that $P_0$ is a binary matrix, obtained from the time-series using the function defined in Eq.~\eqref{eq: burst}.

\textbf{Dimensionality reduction.} The learning process has been modified in order to cope with a large amount of data. We recall that the number of pages has already been reduced by keeping only the ones with bursts of activity. However, it is still a large number of neurons and weights for a Hopfield network. Therefore, instead of training a fully connected network, we reduce the number of links in the following manner. We consider only hyperlink connections between pages and we learn their associated weight. In this way, nodes in the network are linked by human-made connections and it can be seen as a sort of pre-trained, constrained, network. This is a strong assumption as no link in the Hopfield network can be created between pages that are not related by a hyperlink.

\textbf{Hebbian learning.} We use a synaptic-plasticity-inspired computational model~\cite{hebb2005organization} in the proposed graph learning algorithm to compute weights. The main idea is that a co-activation of two neurons results in the enforcement of a connection (synapse) between them. We do not take causality of activations into account. For two neurons $i$ and $j$, their respective activity (number of visits) over time $x_i$ and $x_j$ are compared to determine if they are co-active or not at each time step. We introduce the following similarity measure ${\rm Sim\{i,j,t\}}$ between node $i$ and $j$ at time $t$, where ${\rm Sim\{i,j,t\}}=0$ if $ x_i[t]=x_j[t]=0$ and otherwise:
\begin{equation} \label{similarity}
{\rm Sim}\{i,j,t\} =
\frac{\min(x_i[t], x_j[t])}{\max(x_i[t], x_j[t])}\in[0,1].
\end{equation}
This function compares the ratio of the number of visits, putting more emphasis on the pages receiving a similar amount of visits. 

For each time step $t$, the edge weight $w_{ij}$ between $i$ and $j$ is updated by the following amount: 
\begin{equation} \label{w_update}
\Delta w_{ij} = \begin{cases}
+Sim\{i,j\}, &\text{if $Sim\{i,j\}$ $>$ $\lambda$},\\
0, &\text{otherwise},
\end{cases}
\end{equation}
where $\lambda=0.5$ is a threshold parameter.

The computations are tractable because 1) they are local on the graph, i.e. weight updates depend on a node and its first order neighborhood, 2) weight updates are iterative, and 3) a weight update occurs only between the connected nodes and not among all possible combinations of nodes. These three facts allow us to build a distributed model to speed up computations. For this purpose, we use a graph-parallel Pregel-like abstraction, implemented in the GraphX framework \cite{gonzalez2014graphx}, \cite{xin2013graphx}.

\begin{figure}[t!]
	\centering
	\includegraphics[width=\columnwidth]{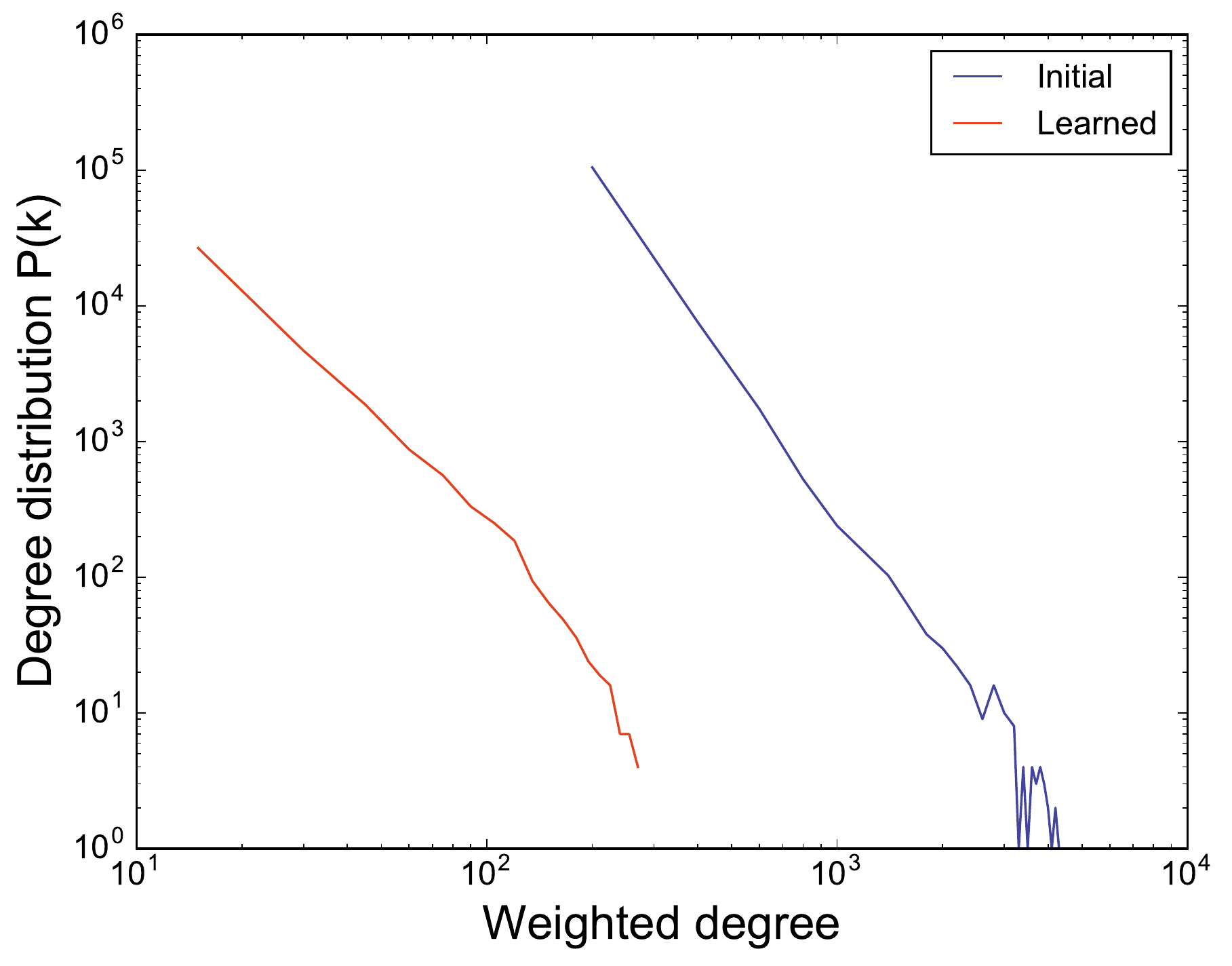}
	\caption{Weighted degree distribution in log-log scale. Linearity in log-log scale corresponds to power-law behavior $P(k) \sim k^{-\gamma}$. Power-law exponent $\gamma=3.81$ for the initial graph (blue), and $\gamma=2.85$ for the learned graph (red). The decrease of the exponent in the learned graph indicates that the weight distribution becomes smoother and the number of greedy hubs drops with the learning.}
	\label{fig: degree_log_log}
\end{figure}

\section{Graph visualization and community detection}\label{visualization}
The structure of each learned network shows patterns corresponding to communities of correlated nodes. We analyze all the monthly Hopfield networks, extract communities and visualize them in the following sections.
We use a heuristic method based on modularity optimization (Louvain method) \cite{blondel2008fast} for community detection. Colors in the graph depict and highlight the detected communities. A resolution parameter controls the size of communities.
To represent the graph in 2D space for visualization, we use a force-directed layout \cite{jacomy2014forceatlas2}. The resulting spatial layout reflects the organization of the network. An interactive version of all visualizations presented in this paper is available online \cite{WikiViz}.

\begin{figure}[t!]
	\centering
	\includegraphics[width=\columnwidth]{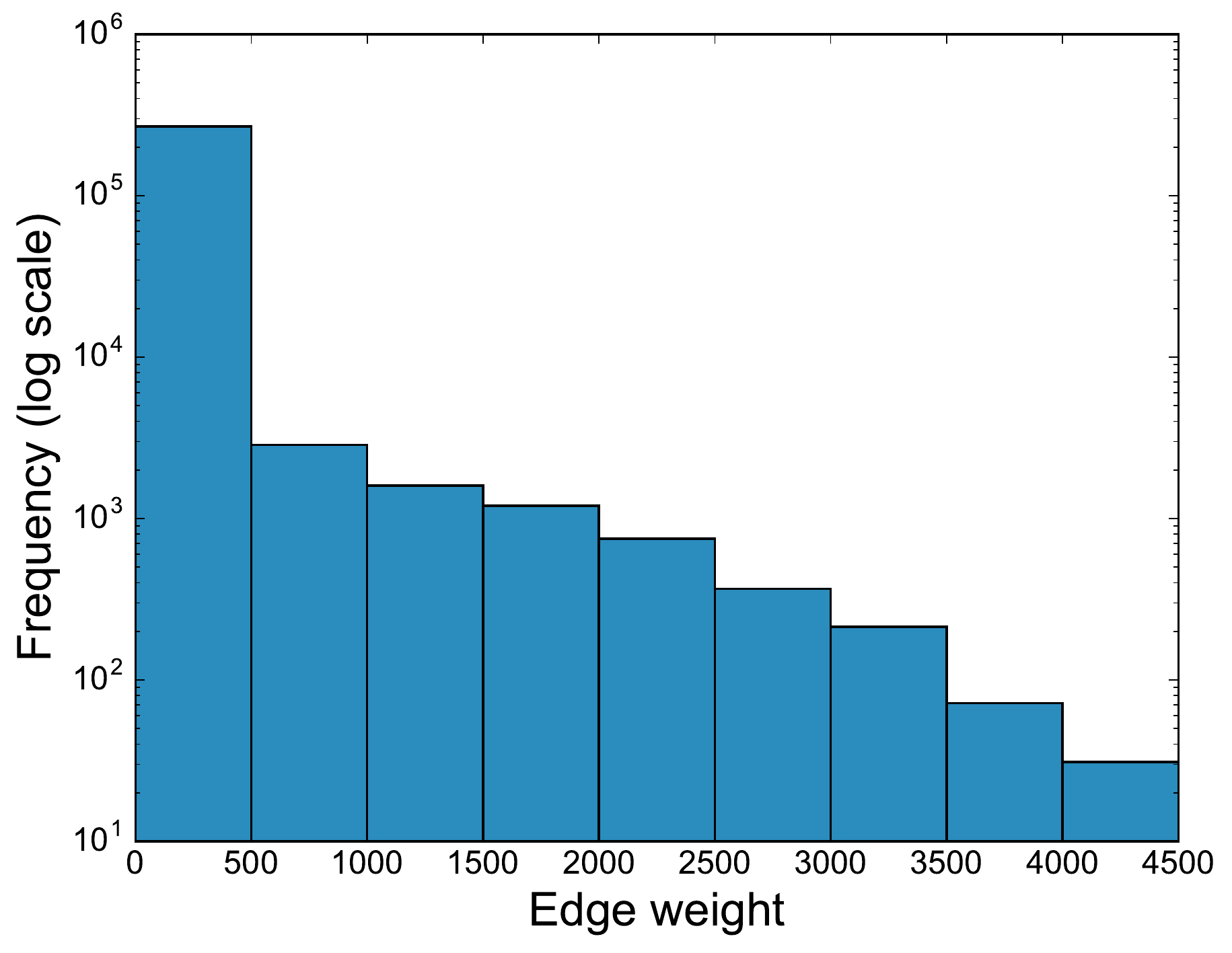}
	\caption{Edge weight distribution in the learned graph in log scale. The learned graph has skewed heavy-tailed weight distribution.}
	\label{fig: edge_weight_distribution}
\end{figure}

\section{Experiments and results}\label{Experimental results}
Using the approach presented in Section~\ref{Graph learning}, we intend to learn collective memories and describe their properties. After the learning, we inspect different resulting networks on different time scales. We extract strongly connected clusters of nodes from these networks and discuss their features. Also, we investigate the temporal behavior of the memories.

{\bf A 7 months network.}
We first learn the graph using the \mbox{7-month} Wikipedia activity dataset. We start from the initial graph of Wikipedia pages connected with hyperlinks, described in Section~\ref{Dataset}. The Hopfield network, the result of the learning process, is a network \mbox{$G = (V,E,W)$}, where $V$ is the set of Wikipedia pages, $E$ is the set of references between the pages, and $W$ is the set of weights, reflecting the similarity of activity patterns between articles.



 

\begin{figure}
	\centering
	\begin{tabular}{cc}
		\includegraphics[width=0.45\columnwidth]{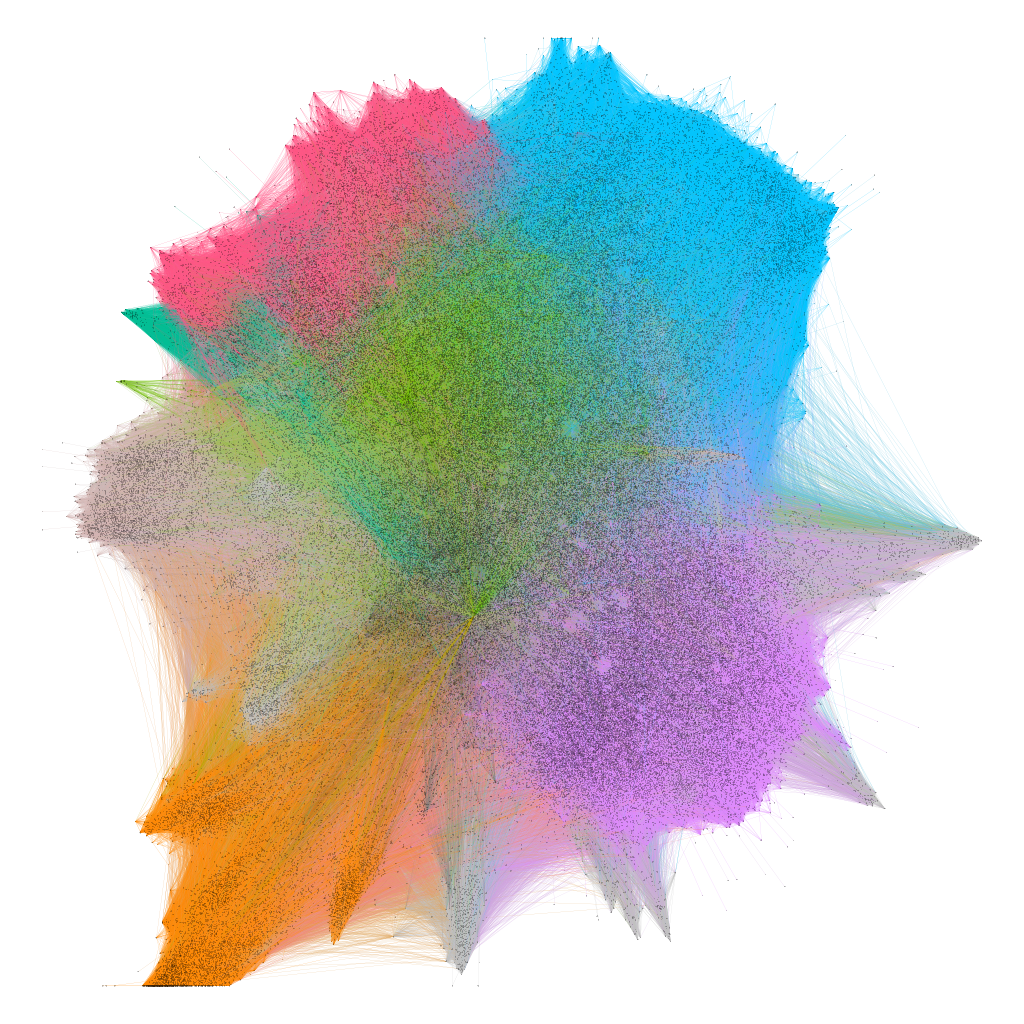} &
		\includegraphics[width=0.45\columnwidth]{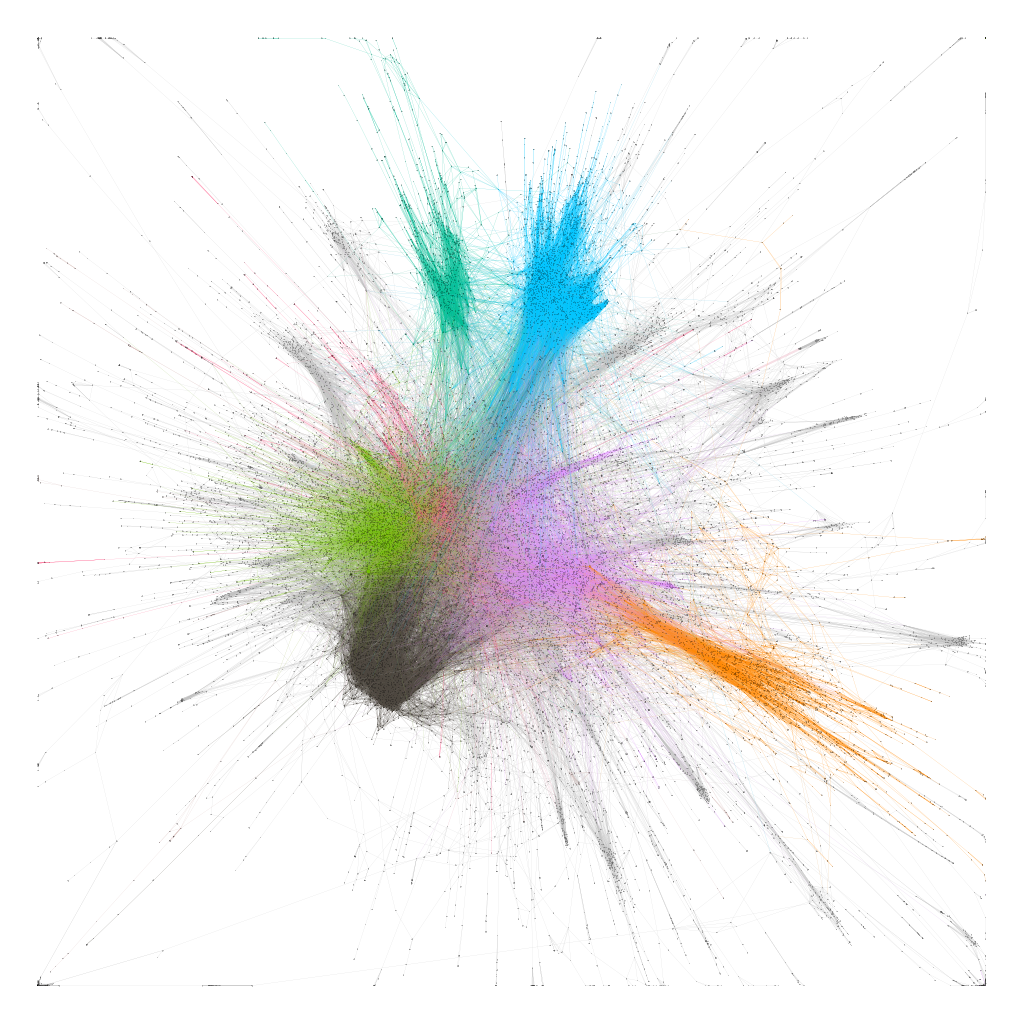}
		\\
		(a) Initial & (b) Learned \\[6pt]
	\end{tabular}
	\caption{Wikipedia graph of hyperlinks (left) and learned Hopfield network (right). Colors correspond to the detected communities. The learned graph is much more modular than the initial one, with a larger number of smaller communities. The layout is force-directed.}
	\label{fig: wiki_graphs}
\end{figure}

After the learning, the majority of weights are 0 (6 297 977, 95.8\%). Only 275 498 edges (4.2\%) have strictly positive weights. For a better visualization, we prune out the edges that have 0-weight. We also delete disconnected nodes that remain after edge pruning and remove small disconnected components of the graph. The number of remaining nodes is 35 839 (31\% of the initial number). Figure~\ref{fig: wiki_graphs} shows snapshots of the initial Wikipedia graph before learning (a) and the learned graph after weight update and pruning (b).

The initial and learned Wikipedia graphs have statistically heterogeneous connectivity (Fig.~\ref{fig: degree_log_log}) and weight patterns (Fig.~\ref{fig: edge_weight_distribution}) that correspond to skewed heavy-tailed distributions.


The degree distribution of the initial graph has a larger power-law exponent $\gamma=3.81$ (Fig.~\ref{fig: degree_log_log}, blue) than the learned one ($\gamma=2.85$). This shows that the initial Wikipedia graph is dominated by large hubs that attract most of the connections to numerous low-degree neighbors. These hubs correspond to general topics in the Wikipedia network. They often link broad topics such as, for instance, the ``United States'' page, with a large number of hyperlinks pointing to it from highly diverse subjects.

\begin{figure}
	\centering
	\includegraphics[width=\columnwidth]{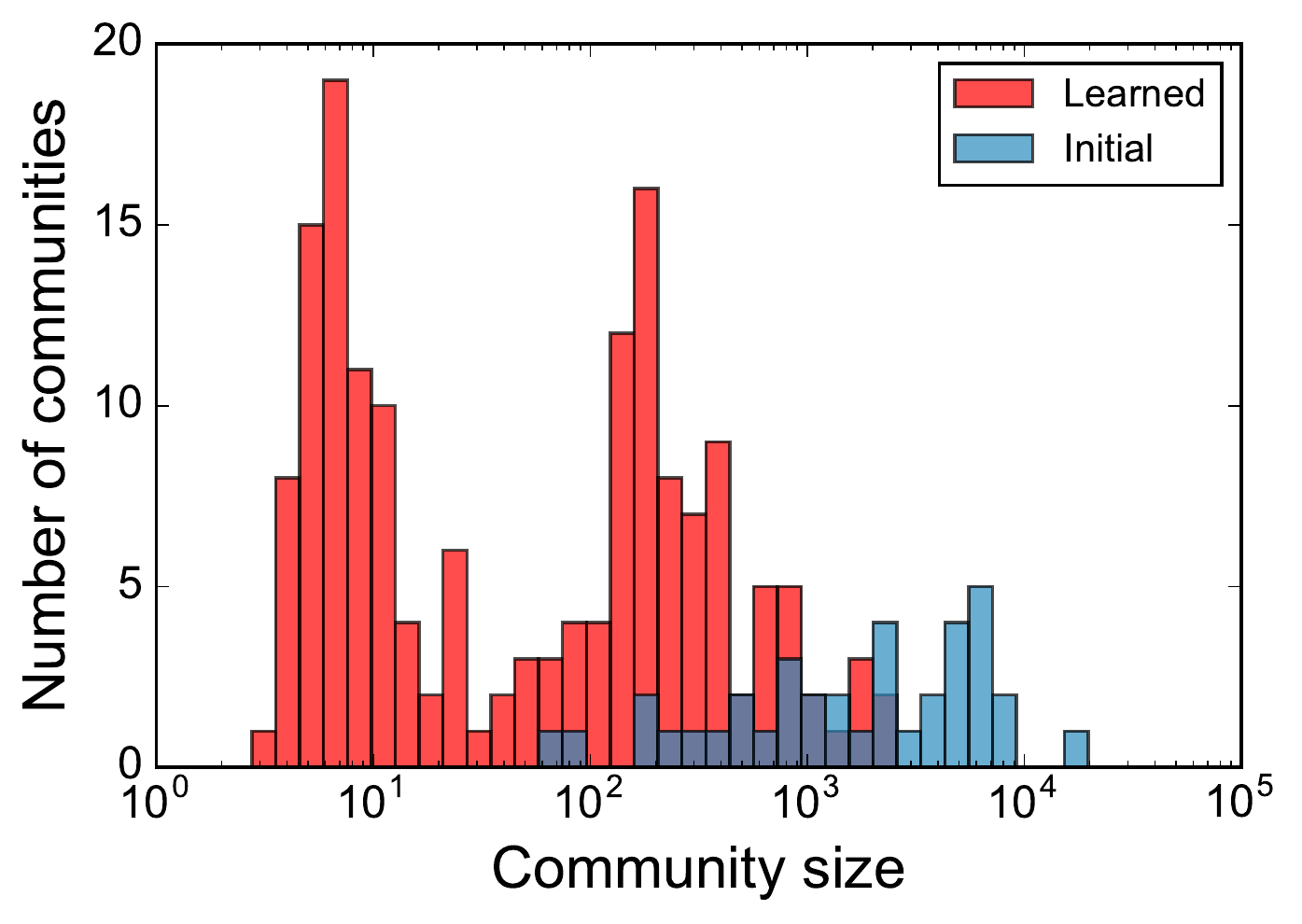}
	\caption{Community size distribution of the initial Wikipedia graph of hyperlinks (blue) and the learned Hopfield network (red). The total number of communities: 32 for the initial graph, 172 for the learned one.}
	\label{fig: modularity}
\end{figure}

We have applied a community detection algorithm, described in Section~\ref{visualization}, to both graphs. The initial Wikipedia graph (Fig.~\ref{fig: wiki_graphs}) is dense and cluttered with a significant number of unused references, while the learned graph reveals smaller and more separated communities. This is confirmed by the community size distribution of the graphs (Fig.~\ref{fig: modularity}). The number of communities and their size change after learning. Initially, the small number of large communities dominate the graph (blue), while after the learning (red) we see a five times increase in the number of communities. Moreover, as a result of the learning, the size of the communities decreases by one order of magnitude. The modularity of the learned graph is 25\% higher, giving the first evidence of the creation of associative structures.

The analysis of each community of nodes in the learned graph gives a glimpse of the events that occurred during the 7-month period. Each cluster is a group of pages related to a common topic such as a championship, a tournament, an awards ceremony, a world-level contest, an attack, an incident, or a popular festive events such as Halloween or Christmas. The graph contains the summary of events that occurred during the period of interest, hence its name of collective memory. 


Before going deeper in the clusters analysis and the information they contain, we investigate the evolution of the graph structure over time.

\textbf{Monthly networks.} To obtain a better, fine-grained view of the collective memories, we focus on a smaller time-scale. Indeed, events attracting the attention of Wikipedia users for a period longer than a week or two are rare. Therefore, we split our dataset into months. Monthly graphs are smaller, compared to the 7-months graph, and contain 10 000 nodes on average. However, the properties and distributions of monthly graphs are similar to the 7-months one, described above. 

\begin{figure*}[t!]
	\includegraphics[width=\textwidth]{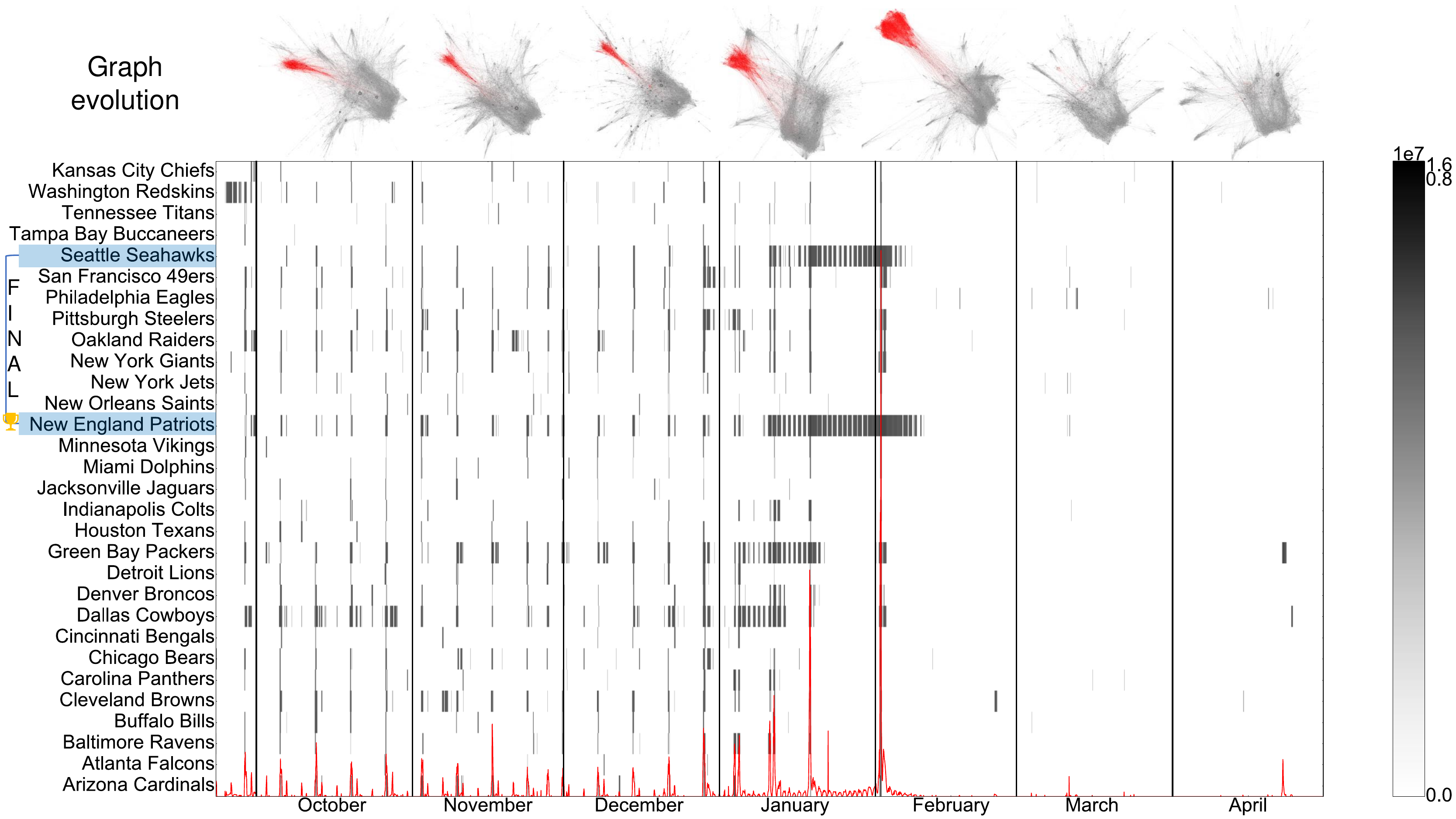}
	\caption{Evolution of the National Football League 2014-2015 championship cluster. We show 30 NFL teams from the main cluster. Top: the monthly learned graph with the NFL cluster highlighted in red. Middle table: visitors activity per hour on the NFL teams' Wikipedia pages in greyscale (the more visits, the darker). Bottom: timeline, in red, of the overall visitor activity in the cluster. }
	\label{fig: SB_evolution}
\end{figure*}

Short-term (monthly) graphs allow to understand and visualize the dynamics of the memory formation in the long-term (7 months) graph. To give an example of the memory evolution, we discuss the cluster of the USA National Football League championship and the collective memories that are mostly related to the previous NFL seasons (Fig.~\ref{fig: SB_evolution}). 
NFL is one of the most popular sports leagues in the USA and its triggers a lot of interest on Wikipedia. Thanks to the high number of visits on this topic we were able to spot a cluster related to the NFL on each of the monthly graphs. Figure~\ref{fig: SB_evolution} shows information about the NFL clusters. The top part of the figure contains the learned graphs, for each month, where the NFL cluster is highlighted in red.
The final game of the 2014 season, Super Bowl XLIX, had been played on February 1, 2015. This explains the increase in size until February where it reaches its maximum. The activity collapses after this event and the cluster disappears.
For the sake of interpretability, we extracted 30 NFL team pages from the original cluster (485 pages) to show the details of the evolution in time as a table on Fig.~\ref{fig: SB_evolution}. This fraction of the nodes reflects overall dynamics in the entire cluster. Each row describes the hourly activity of a page, while the columns split the plot into months.
The sum of visits for the selected pages is plotted as a red line on the bottom.
The dynamics of the detected cluster reflects the real timeline of the NFL championship. The spiking nature of the overall activity corresponds to weekends when most of the games were played. Closer to the end of the championship, the peaks become stronger, following the increasing interest of fans. We see the highest splash of the activity on 1 February, when the final game was played.

We want to emphasize that this cluster, as well as all the others, was obtained in a completely unsupervised manner. Football team pages were automatically connected together in a cluster having ``NFL'' as a common topic. Moreover, the cluster is not formed by one Wikipedia page and its direct neighbors, it involves many pages with distances of several hops on the graph. 


The NFL championship case is an example of a periodic (yearly) collective memory. The interest increases over the months until the expected final event. Accidents and incidents are different types of events as they appear suddenly, without prior activity. Our proposed learning method allows for the detection of these kinds of collective memories as well. We provide examples of three accidents to demonstrate the behavior of the collective memory in case of an unexpected event.

\begin{figure*}[!t]
	\begin{tabular}{ccc}
		\includegraphics[width=0.315\textwidth, trim={0cm 8.5cm 0cm 8.5cm}, clip]{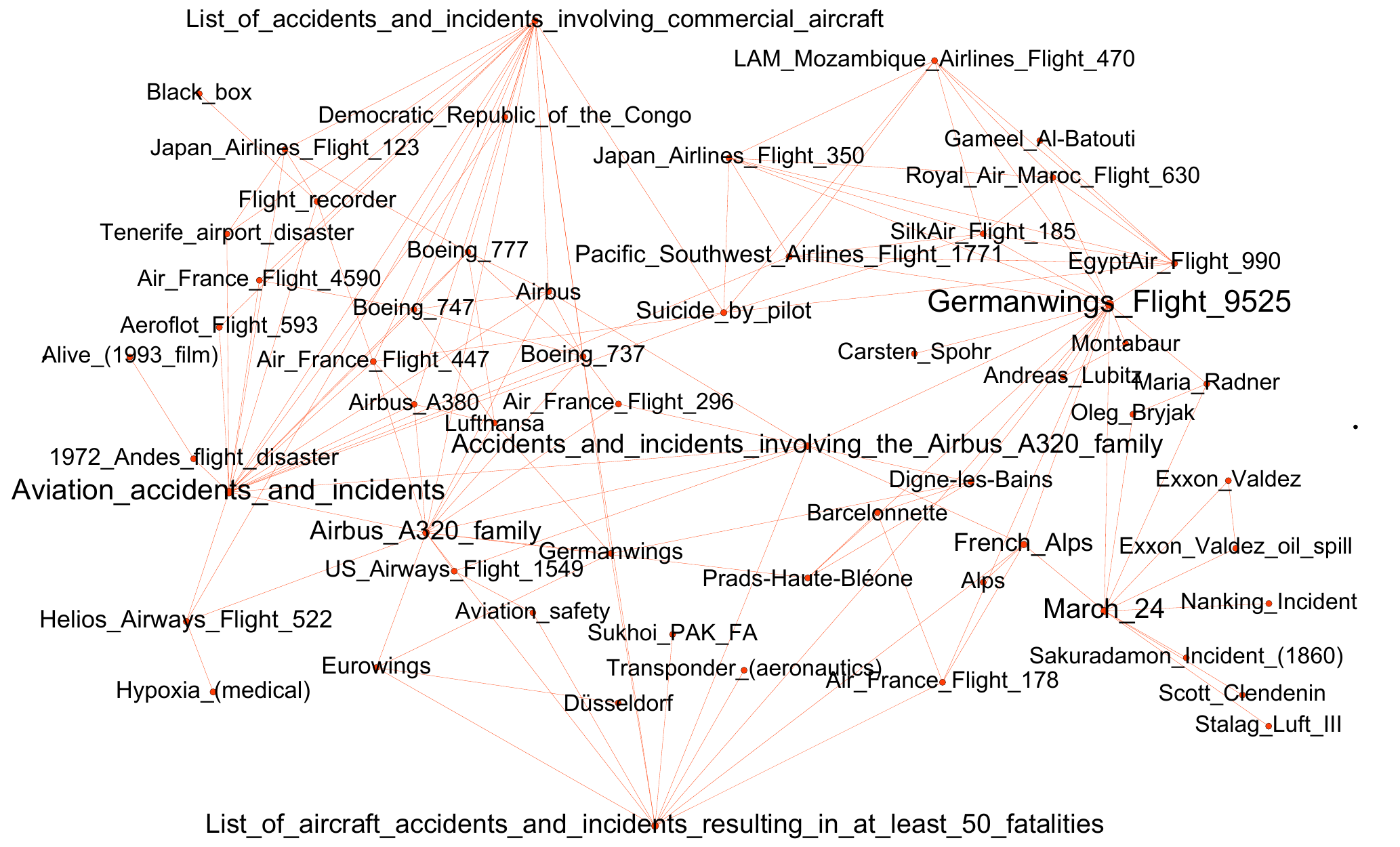} &
		\includegraphics[width=0.315\textwidth, trim={0cm 8.5cm 0cm 8cm}, clip] {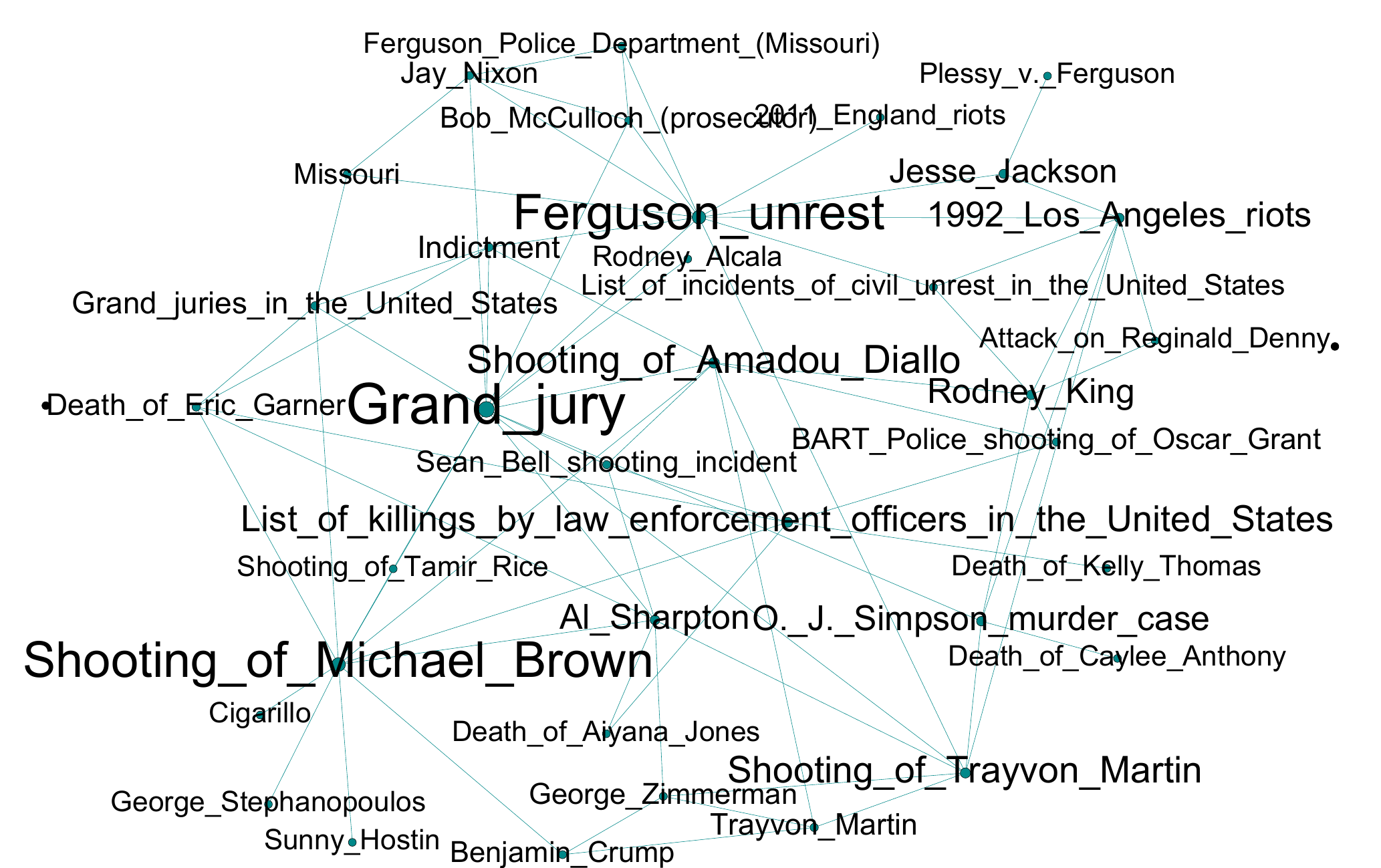} &
		\includegraphics[width=0.315\textwidth, trim={0cm 8.5cm 0cm 8.5cm}, clip]{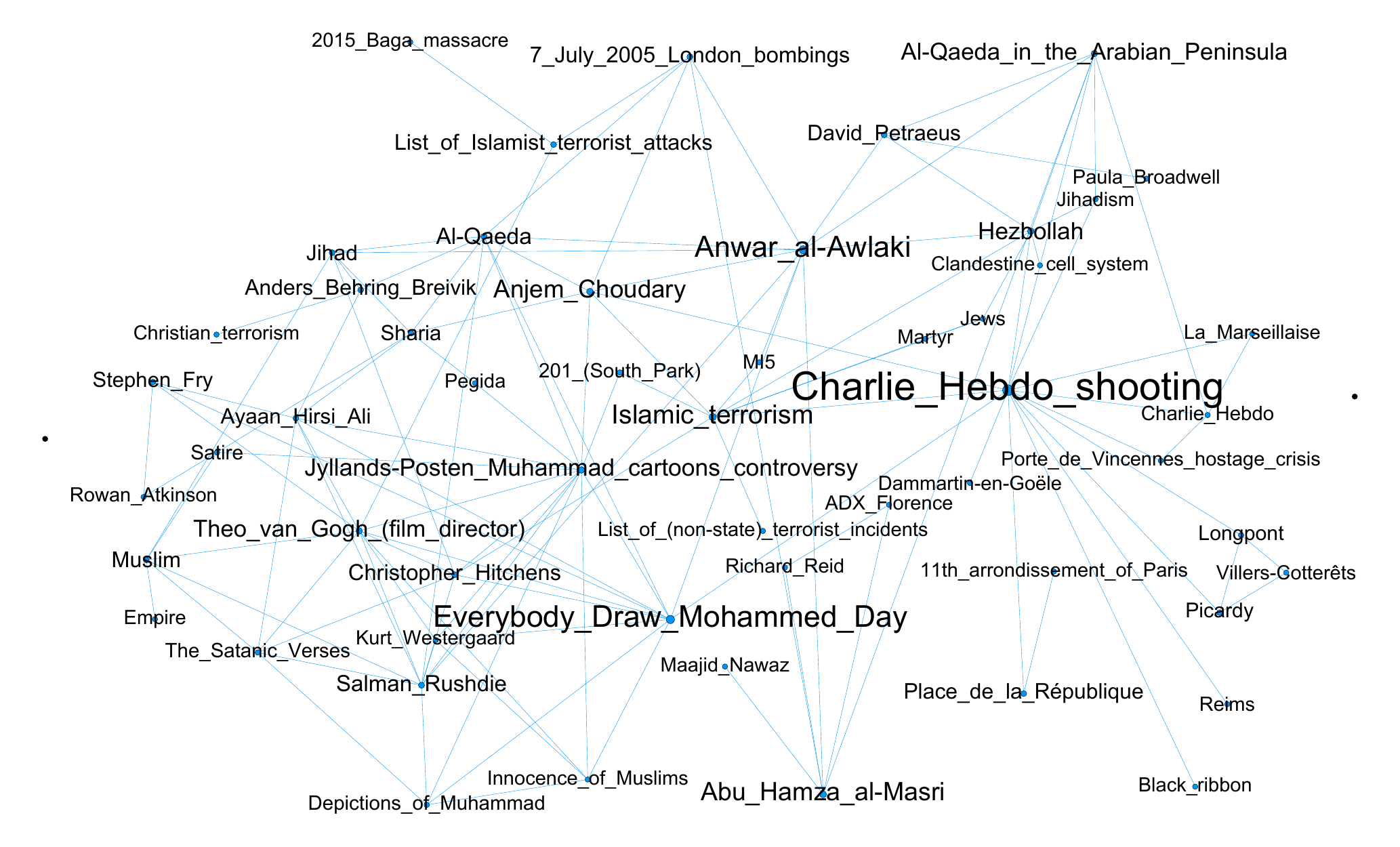}\\
		\includegraphics[width=0.315\textwidth,  trim={5cm 0cm 5cm 0cm}, clip] {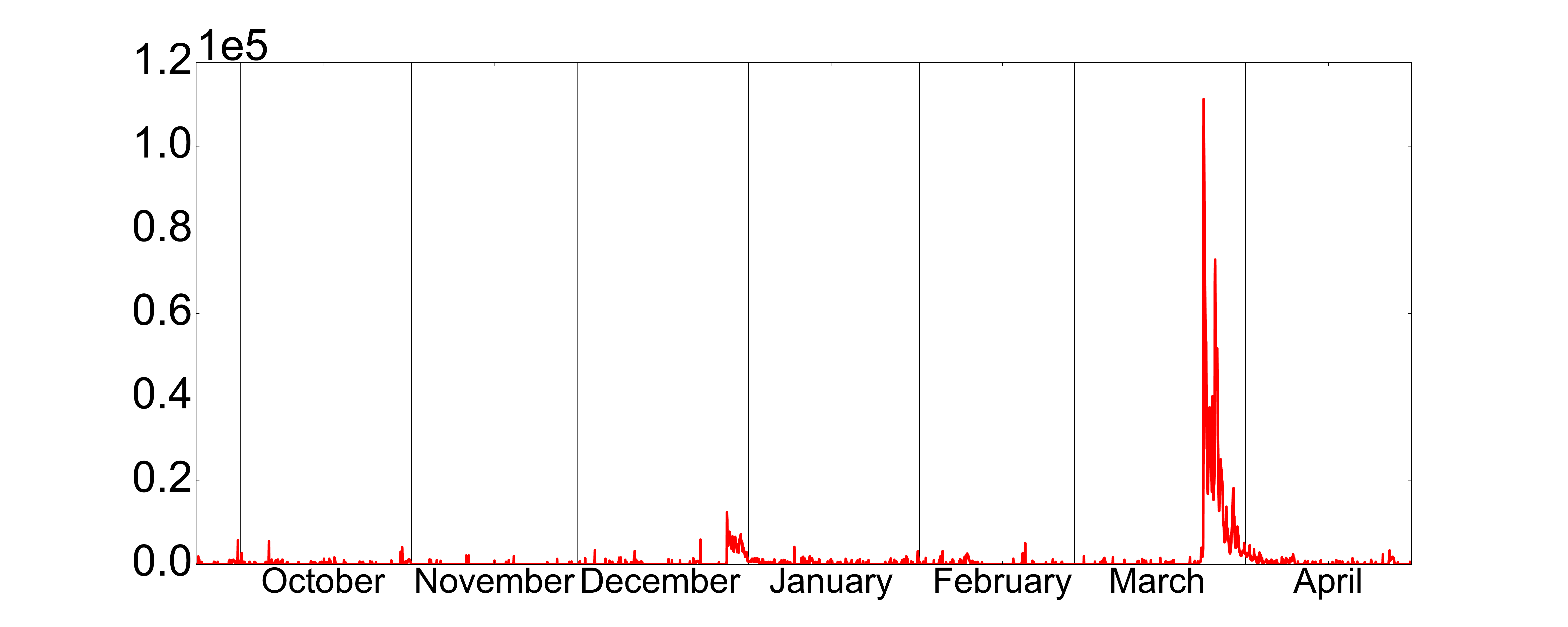} &
		\includegraphics[width=0.315\textwidth,  trim={5cm 0cm 5cm 0cm}, clip] {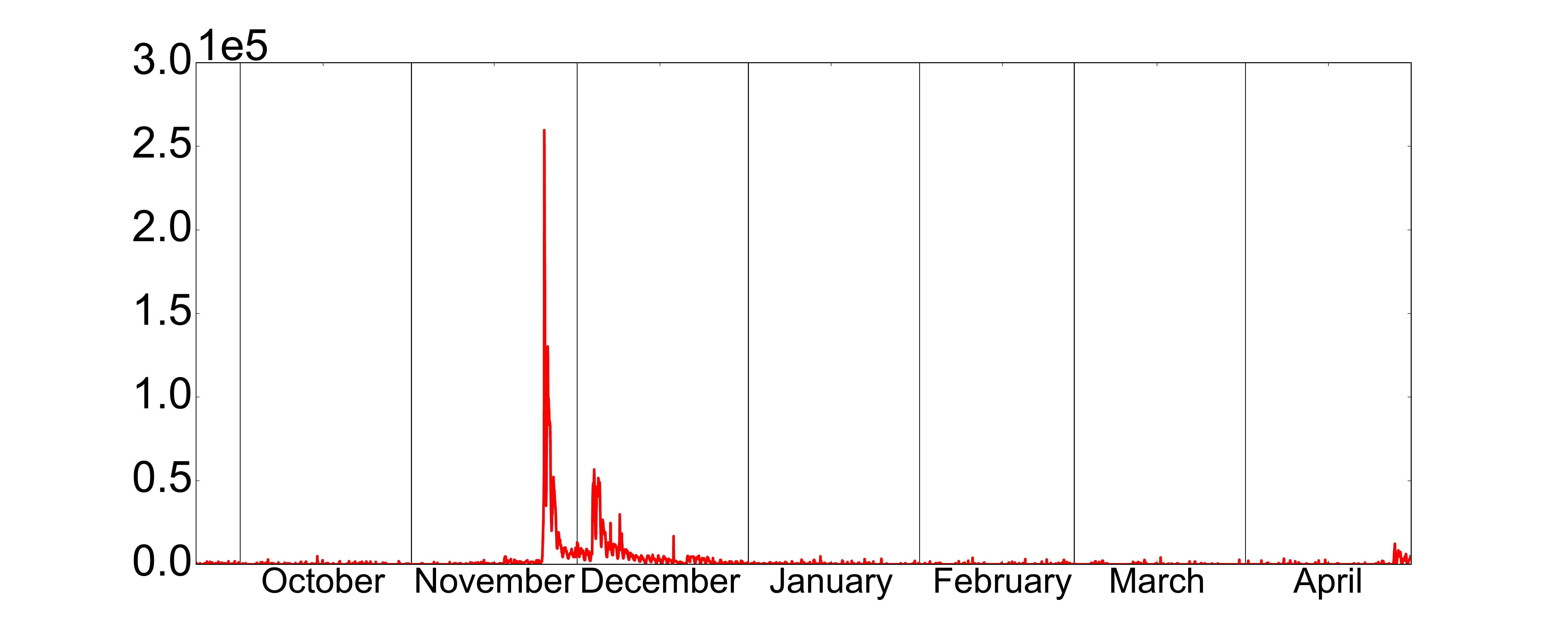} &
		\includegraphics[width=0.315\textwidth, trim={5cm 0cm 5cm 0cm}, clip] {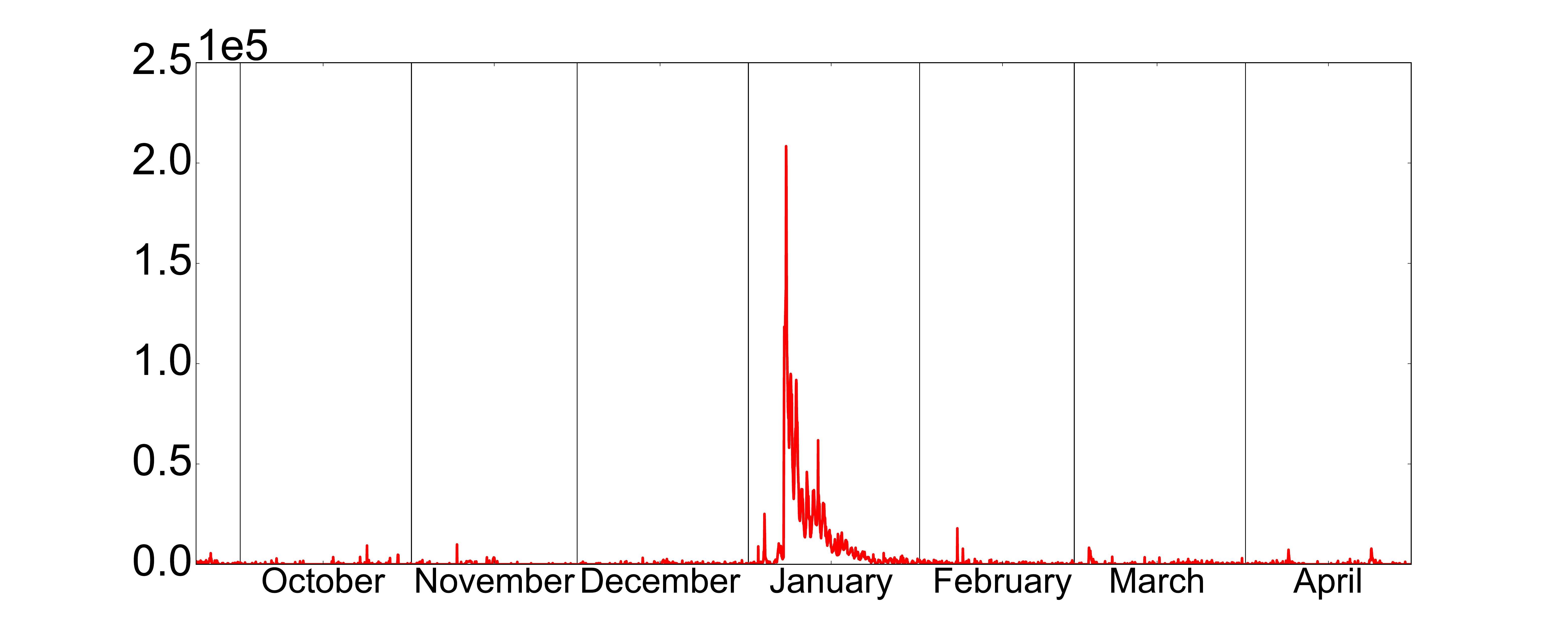} \\
		(a) Germanwings 9525 crash & (b) Ferguson unrest & (c) Charlie Hebdo attack
	\end{tabular}
	\caption{Collective memory clusters (top) and overall activity of visitors in the clusters (bottom).}
	\label{fig: dynamics}
\end{figure*}

We pick three core events among 172 detected and discuss them to show the details of memory detection by our approach. Figure~\ref{fig: dynamics} shows the extracted clusters from the learned graph (top) and the overall timeline of the clusters' activity (bottom).

\textit{Charlie Hebdo shooting.} 7 January 2015. This terrorist attack is an example of an unexpected event. The cluster emerged over a period of 72 hours, following the attack. All pages in the cluster are related to the core event. Strikingly, a look at the title of the pages is sufficient to get a precise summary of what the event is about. There is a sharp peak of activity on the first day of the attack, slowly decaying over the following week.

\textit{Germanwings flight 9525 crash.} 24 March 2015. This cluster not only involves pages describing the crash or providing more information about it, but also the pages of similar events that happened in the past. It includes, for example, a page enumerating airlane crashes and the page of a previous crash that happened in December 2014, the Indonesia AirAsia Flight 8501 crash. As a result, the memory of the event is connected to the memory of the Flight 8501 crash, that is why we can see an increase of visits in December. This is an example where our associative memory approach not only connects pages together but also events.

\textit{Ferguson unrest. Second wave.} November 24, 2014 -- December 2, 2014. This is an example of an event that has official beginning and end dates. A sharp increase in the activity at the beginning of protests highlights the main event. This moment triggers the core cluster emergence. We also see that the cluster becomes active once again at the end of the unrest.


Eventually, we conclude our exploration of the clusters of the learned graphs by providing on
Table \ref{table: collective_memories} a list of handpicked page titles inside each cluster that refer to previous events and related subjects. The connected events occurred outside of the 7-months period we analyze. This illustrates the associative features of our method. Firstly, pages are grouped by events with the help of visitors activity. Secondly, events themselves are connected together by this activity. Memory is made of groups of concepts tightly connected and these groups are in turn connected together through concepts they share.

\begin{table*} [!t]
	\caption{Collective memories triggered by core events.}
	\label{table: collective_memories}
	\begin{minipage}{\textwidth}
		\begin{center}
			\begin{tabular}{ lll }
				\toprule
				\textbf{Charlie Hebdo attack} & \textbf{Germanwings 9525 crash} & \textbf{Ferguson unrest} \\
				\midrule
				Porte de Vincennes hostage crisis & Inex-Adria Aviopromet Flight 1308 & Shooting of Tamir Rice \\
				Al-Qaeda & Pacific Southwest Airlines Flight 1771 & Shooting of Amadou Diallo \\
				Islamic terrorism & SilkAir Flight 185 & Sean Bell Shooting Incident \\ 
				Hezbollah & Suicide by pilot & Shooting of Oscar Grant \\
				2005 London bombings & Aviation safety & 1992 Los Angeles riots \\ 
				Anders Behring Breivik & Air France Flight 296 & O.J. Simpson murder case \\ 
				Jihadism & Air France Flight 447 & Shooting of Trayvon Martin \\ 
				2015 Baga massacre & Airbus & Attack on Reginald Denny \\
				\bottomrule
			\end{tabular}
		\end{center}
		\bigskip
	\end{minipage}
\end{table*}

{\bf Recalling memories.}
In this section, our goal is to test the hypothesis that the proposed method, as a memory, allows recalling events from partial information. We emulate recall processes using the Hopfield network approach, described in Section~\ref{Graph learning}. We show that the learned graph structure can recover a memory (cluster of pages and its activations) from an incomplete input. 


We create incomplete patterns for the Hopfield network by selecting randomly a few pages contained in a chosen cluster. We built the input matrix $P_0$ setting to $(-1)$ (inactive state) all the time-series except for the few pages selected. We then apply iteratively Eq.~\eqref{eq: hopfield}.

\begin{figure*} [!t]
	\begin{tabular}{c}
		\includegraphics[width=\textwidth, trim={0cm 0cm 0cm 0cm}, clip]{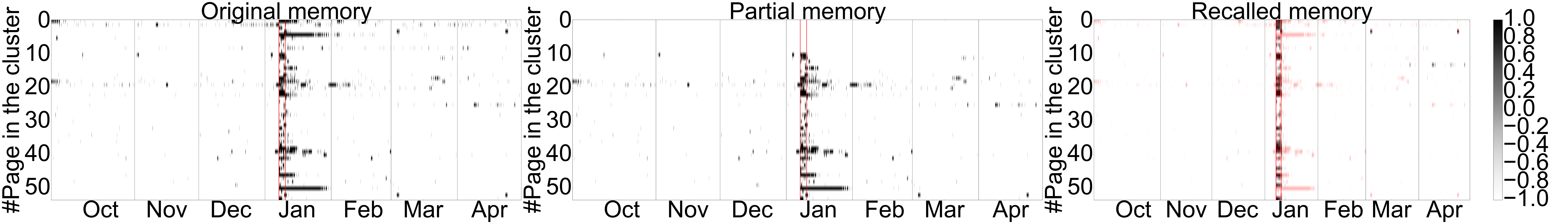}
	\end{tabular}
	\caption{Recall of an event from a partial pattern (Charlie Hebdo attack). The red vertical lines define the start of the event and its most active part, ending 72 hours from the start. Left: full activity over time of the pages in the cluster. Middle: pattern with 20\% of it set inactive. Right: result of the recall using the Hopfield network model of associative memory. In light red are shown the difference with the original pattern (the forgotten activity).}
	\label{fig: remind}
\end{figure*}

The results of the experiment are illustrated with an example on Figure \ref{fig: remind}. We selected the cluster associated to the Charlie Hebdo Attack. From the list of pages, we select a subset of it, here $80\%$. We applied the learned graph for the month of January, when the memory was detected. After the recall (Fig. 8, right), most of the cluster is recovered at the correct position in time. Note that the model forgets a part of the activity, plotted in light red. This missing part is made of pages that are active outside of the time of the event, giving the evidence that they are not directly related (or weakly related) to the event.

To evaluate the performance of the remembering, we mute different fractions of nodes in memory clusters and compute recall errors. We define the recall accuracy $A$ as the ratio of correctly recalled activations over the number of initial activations. The error is given by $1-A$. Figure \ref{fig: evaluation} shows the results of the evaluation. We consider three types of errors. First, we measure the accuracy of a recalled pattern over 7 months (green). Second, the accuracy of a recalled pattern over a period of 72 hours after the start of the event (blue). And third, a relaxed accuracy of the recall during the same period of time, with a correct recovery if an initially active node is active at least once in the recalled pattern (red). 
As expected, the quality of the recalled memory increases with the quality of the input pattern. The better performance of the recall in the 72 hours zone is due to the memory effect that focuses on the most active part of the cluster and forgets small isolated activity spikes of the individual pages scattered over the 7 months.

\begin{figure*} [!t]
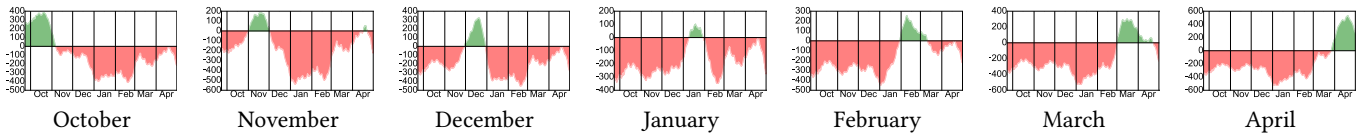

	\begin{tabular}{ccccccc}
		\includegraphics[width=0.127\textwidth, trim={10cm 2cm 44cm 2cm}, clip]{figures/oct_recall.pdf} &
		\includegraphics[width=0.127\textwidth, trim={10cm 2cm 44cm 2cm}, clip]{figures/nov_recall.pdf} &
		\includegraphics[width=0.127\textwidth, trim={10cm 2cm 44cm 2cm}, clip]{figures/dec_recall.pdf} &
		\includegraphics[width=0.127\textwidth, trim={10cm 2cm 44cm 2cm}, clip]{figures/jan_recall.pdf} &
		\includegraphics[width=0.127\textwidth, trim={10cm 2cm 44cm 2cm}, clip]{figures/feb_recall.pdf} &
		\includegraphics[width=0.127\textwidth, trim={10cm 2cm 44cm 2cm}, clip]{figures/mar_recall.pdf} &
		\includegraphics[width=0.127\textwidth, trim={10cm 2cm 44cm 2cm}, clip]{figures/apr_recall.pdf}\\
		October &
		November &
		December &
		January &
		February &
		March &
		April
	\end{tabular}
	\caption{Recalled activity over time for each monthly learned graph.}
	\label{fig: recall}
\end{figure*}





In Figure \ref{fig: recall}, we show the results of the recall process for each monthly learned graph when the entire time-series are given at the input. The output of the graph is summed up to create a global curve of activity over the 7 months. The initial global activity is subtracted from the output curve to obtain the final curve plotted on the figure. Red areas on the plot correspond to a low activity in the output, hence an absence of memory in the zone. Green areas represent positive recalls. The memory of each of the monthly graph corresponds to its learning period. This demonstrates the global effectiveness of the learning and the recall process.

\begin{figure}[t!]
	\centering
	\includegraphics[width=\columnwidth]{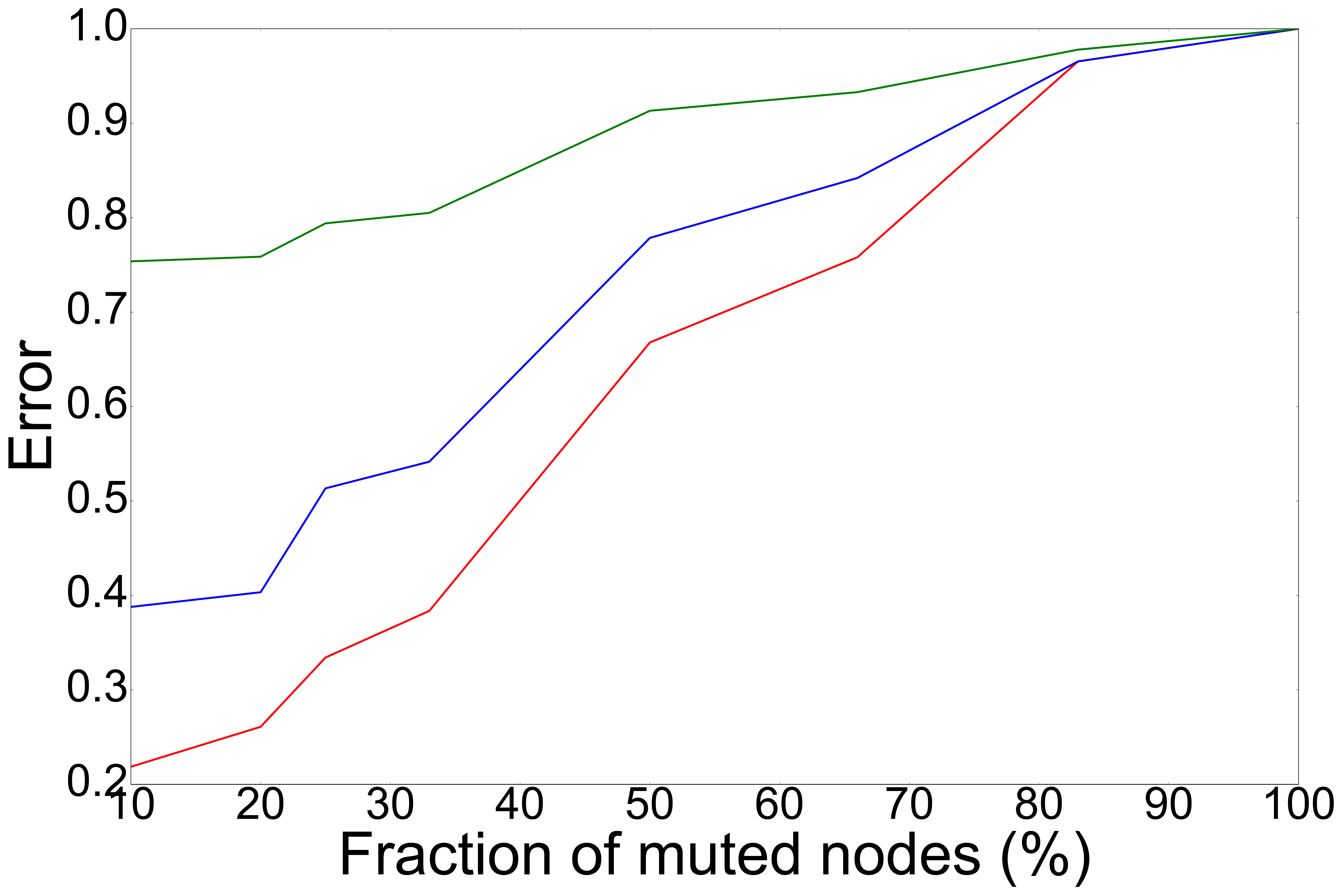}
	\caption{Plot of the recall error rate with three different error measures.}
	\label{fig: evaluation}
\end{figure}

\section{Conclusions and future work}\label{Conclusions}
In this paper, we propose a new method that allows learning and remembering collective memories in an unsupervised manner. To extract memories, the method analyses the Wikipedia Web network and hourly viewership history of its articles. 
The collective memories are summaries of the events that raised the interest of Wikipedia visitors. The approach is able to perform an efficient knowledge retrieval from the Wikipedia encyclopedia allowing to highlight the changing interests of visitors over time and to shed some light on their collective behavior.

This approach is able to handle large-scale datasets. We have noted experimentally a high robustness of the method to the tuning of the parameters. This will be investigated further in future work.



This work opens new avenues for dynamic graph-structured data analysis. For example, the proposed approach could be used in a framework for automated event detection, monitoring, and filtering in network structured visitor activity streams.
The resulting framework is also of interest for the understanding of the human memory and the simulation of an artificial memory.



\section{Tools, implementation, code and online visualizations} \label{reproducible}

All learned graphs (overall September-April, monthly activity, and localized events) are available online~\cite{WikiViz}
to foster further exploration and analysis. For graph visualization we used the open source software package Gephi~\cite{bastian2009gephi} and layout ForceAtlas2~\cite{jacomy2014forceatlas2}. We used Apache Spark GraphX~\cite{xin2013graphx}, \cite{gonzalez2014graphx} for graph learning implementation and graph analysis. The presented results can be reproduced using the code for Wikipedia dynamic graph learning, written in Scala~\cite{Miz2017}. The dataset is available on Zenodo~\cite{benzi_kirell_2017_886484}.

\begin{acks}
	
	We would like to thank Micha\"{e}l Defferrard and Andreas Loukas for fruitful discussions and useful suggestions.
	
	The research leading to these results has received funding from the European Union's H2020 Framework Programme (H2020-MSCA-ITN-2014) under grant agreement n\textsuperscript{o}~642685 MacSeNet.
	
\end{acks}